\documentclass[preprint,showpacs,prd,tightenlines]{revtex4}
\usepackage{graphicx}
\usepackage{slashed}

\if@mathematic
   \def\vec#1{\ensuremath{\mathchoice
                     {\mbox{\boldmath$\displaystyle\mathbf{#1}$}}
                     {\mbox{\boldmath$\textstyle\mathbf{#1}$}}
                     {\mbox{\boldmath$\scriptstyle\mathbf{#1}$}}
                     {\mbox{\boldmath$\scriptscriptstyle\mathbf{#1}$}}}}
\else
   \def\vec#1{\ensuremath{\mathchoice
                     {\mbox{\boldmath$\displaystyle#1$}}
                     {\mbox{\boldmath$\textstyle#1$}}
                     {\mbox{\boldmath$\scriptstyle#1$}}
                     {\mbox{\boldmath$\scriptscriptstyle#1$}}}}
\fi
\newcommand{\intd}{\mathrm{d}}

\newcommand{\gev}{\mathrm{GeV}}
\newcommand{\pp}{\pi^+\pi^-}
\newcommand{\kk}{K^+K^-}
\newcommand{\jpsi}{J/\psi}
\newcommand{\psp}{\psi(2S)}
\newcommand{\ppjpsi}{\pp J/\psi}
\newcommand{\pppsp}{\pp \psp}
\newcommand{\EE}{e^+e^-}

\begin{document}

\title{Cross sections of $e^+e^-\to \gamma VV$
and $e^+e^-\to \gamma \gamma V$}

\author{K.~Zhu}
\email{zhuk@ihep.ac.cn}

\author{C.~Z.~Yuan}
\email{yuancz@ihep.ac.cn}

\author{R.~G.~Ping}
\email{pingrg@ihep.ac.cn}

\affiliation{Institute of High Energy Physics, Chinese Academy of
Sciences, P.O. Box 918, Beijing 100049, China}

\date{\today}

\begin{abstract}

Cross sections of the $e^+e^-$ annihilation into one photon plus two
vector mesons, such as $J/\psi \rho$, $J/\psi \phi$, and $\rho\phi$
at the center of mass energy $\sqrt{s}=10.58~\gev$ are calculated.
These shed light on the measurement of the $e^+e^-$ annihilation
cross section via radiative return in $B$-factories. At another
center of mass energy $\sqrt{s}=3.097~\gev$, namely the $\jpsi$
resonance peak, the processes $e^+e^-\to\gamma VV$ and $e^+e^-\to
\gamma\gamma V$ (with $V$ being $\rho$, $\omega$, or $\phi$) are
calculated with similar method. These calculations give estimation
of the background levels in the study of radiative or double
radiative decays of $\jpsi$.

\end{abstract}


\maketitle

\section{Introduction}

The measurement of the $\EE$ annihilation cross section via
initial state radiation ($ISR$) turns out to be very fruitful at
the $B$-factories due to the large data samples collected on the
$\Upsilon(4S)$ resonance. The study of the final states with a
charmonium ($\jpsi$ or $\psp$) and light hadrons such as $\pp$ and
$\kk$ results in the discovery of many unexpected charmonium-like
states, such as the $Y(4008)$, $Y(4260)$, $Y(4360)$, and
$Y(4660)$~\cite{Aubert:2005rm,Aubert:2006ge,Yuan:2007sj,Wang:2007ea}.
But one should notice that, $e^+e^- \to \gamma \rho J/\psi$ is
actually an important background in the measurement of the
$\pi^+\pi^-J/\psi$ production via $ISR$
processes~\cite{Aubert:2005rm,Yuan:2007sj}; and this is also true
in the case when final state $\pi^+\pi^-\psi(2S)$ was
concerned~\cite{Aubert:2006ge,Wang:2007ea}. Another similar case
is in the study of $K^+K^-J/\psi$ via $ISR$, here background from
$\gamma \phi J/\psi$ should be considered. Actually, of a recent
observation at Belle~\cite{Yuan:2007bt}, this background has been
estimated and subtracted by a requirement on the $K^+K^-$
invariant mass does not agree with a $\phi$. While the subtraction
of $\gamma \phi J/\psi$ background is possible due to the narrow
width of the $\phi$ meson, it is almost impossible for the
$\ppjpsi$ and $\pppsp$ case, where the $\pp$ invariant mass
distribution is not very different from the wide $\rho$ resonance,
so a calculation of the cross section is desired.

At lower energies, in the studies of $J/\psi \to \gamma VV$ at
BES~\cite{Xu:2007ca, Ablikim:2006dw, Ablikim:2006ca}, where $V$
denotes light neutral vector meson such as $\rho$ (or $\omega$,
$\phi$), direct productions $e^+e^-\to \gamma VV$ are backgrounds
too. Analogously, $e^+e^-\to \gamma \gamma V$ are backgrounds in
the $J/\psi$ double radiative decay channels~\cite{Xu:2004qj} as
well as in the energy region of few tens of $\mathrm{MeV}$ around
$\phi$ resonance where processes $e^+e^-\to \omega\pi^0$ and
$e^+e^-\to \rho\pi^0$ are measured at
KLOE~\cite{Ambrosino:2006hb,Ambrosino:2007wf}. We can see that,
although the direct production of $e^+e^-\to \gamma VV$ or
$e^+e^-\to \gamma \gamma V$ is believed only at order
$\mathcal{O}(\alpha^3)$, the high luminosity at the $\phi$,
$B$-factories and the forthcoming BESIII will provide
opportunities to explore these rare processes.

The process $e^+e^-\to $ hadrons at center of mass (CM) energy
$\sqrt{s}$ far below the $\mathrm{Z}^0$ mass is dominated by
annihilation via a single virtual photon with charge-conjugation
parity $C=-1$. Recently BaBar~\cite{Aubert:2006we} presented the
first observation of the exclusive reactions $e^+e^-\to
\rho^0\rho^0$ and $e^+e^-\to \phi\rho^0$, in which the final
states are even under charge conjugate, and therefore cannot be
produced via just a single photon. A possible interpretation is
that these $C=+1$ final states are produced in the
two-virtual-photon annihilation processes, i.e., they arise from
$e^+e^-$ annihilation into two virtual photons with each virtual
photon converts into a vector meson. The rates predicted by the
above mechanism can be computed unambiguously using the effective
vector meson-photon couplings determined from the leptonic widths
of the mesons. Based on this assumption, the authors of
Ref.~\cite{Davier:2006fu} calculated the cross sections of series
of these processes and the results are in good agreement with
BaBar's measurements.

In this article, we calculate the cross sections of several
$e^+e^-\to \gamma VV$ processes, such as $e^+e^-$ annihilate into
one photon plus $J/\psi \rho$, $J/\psi \phi$ and $\rho\phi$ at the
CM energy $\sqrt{s}=10.58~\gev$ using the same method proposed in
Ref.~\cite{Davier:2006fu}. We also calculate $e^+e^-$ annihilate
into one photon plus $VV$, where $VV$ denotes $\omega\phi$ or
$\rho\phi$ at $\sqrt{s}=3.097~\gev$, in addition to calculate
$e^+e^-$ annihilate into two photons plus a $\rho$, a $\phi$ or an
$\omega$. The structure of this paper is, after a short
introduction of the model, we give some details of how to
calculate the amplitude of $e^+e^-\to (n)\gamma(m)\gamma^*$, then
the numerical results and finally a brief discussion.

\section{Amplitude of $e^+e^-\to (n)\gamma(m)\gamma^*$}
Our motivation focuses on investigating the non-resonance
contributions to the processes of $e^+e^-\to \gamma V_1V_2$ and
$\gamma\gamma V$, where the vector mesons of $V_1$ and $V_2$ have
different quark contents. We assume the generic reaction $e^+e^-\to
\gamma V_1V_2$ proceeds via an intermediate $e^+e^-\to \gamma
\gamma^*\gamma^*$ process, and the two virtual photons converting
into $V_1$ and $V_2$ with effective couplings $e/f_1$ and $e/f_2$
respectively. Then the differential cross section simply reads as
\begin{equation}
\intd\sigma_{\gamma V_1V_2}=\left(\frac{e}{f_1}\right)^2
\left(\frac{e}{f_2}\right)^2 \intd\sigma_{\gamma
  \gamma^*_1\gamma^*_2}(m^2_{V_1},m^2_{V_2})\;,
\label{eq:two-photon-model}
\end{equation}
where $\intd\sigma_{\gamma\gamma^*_1\gamma^*_2}
(m^2_{V_1},m^2_{V_2})$ is the differential cross section of
$e^+e^-\to \gamma \gamma^*\gamma^*$, which depends on the virtual
photon masses, i.e. the masses of final vector mesons. The
effective photon-meson couplings can be directly defined using the
leptonic widths of the vectors
\begin{equation}
\Gamma^V_{ee}=\frac{\alpha}{3}\left(\frac{e}{f_V}\right)^2 m_V \;,
\label{eq:leptonic-width}
\end{equation}
when narrow widths approximation is used. Notice that the $\rho$
meson cannot be described as a narrow vector meson properly for
its somewhat large width, a complete consideration should take an
integral over its mass distribution. But from a practical point of
view the narrow width approximation works well, and the $\rho$
meson mass distribution only contribute a correction less than
$8\%$~\cite{Davier:2006fu}. So we will adopt narrow widths
approximation in following computation. For another similar
reaction $e^+e^-\to \gamma \gamma V$, the cross section formula is
similar to Eq.~(\ref{eq:two-photon-model}) except one virtual
photon be replaced by a real one as long as the interference
between the two real photons can be neglected.

\begin{figure}[htb]
\centering
\includegraphics[width=0.8\textwidth]{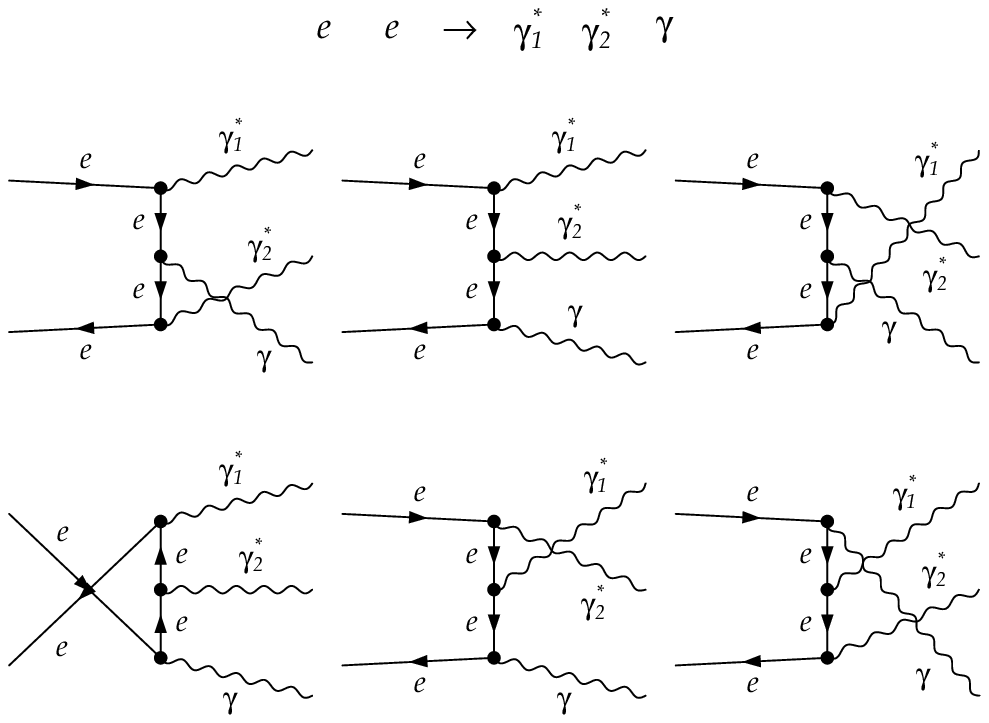}
\caption{Feynman diagrams of $e^+e^-\to \gamma\gamma^*\gamma^*$ at
tree order.} \label{fig:twophotonew}
\end{figure}

In order to calculate the cross sections of $e^+e^-\to \gamma VV$
or $e^+e^-\to \gamma \gamma V$, from
Eq.~(\ref{eq:two-photon-model}) it's clear that an essential work
is to calculate the cross section of a pure QED process, which can
be represented by a more general form $e^+e^-\to
(n)\gamma(m)\gamma^*$, i.e. an electron and a positron annihilate
into $n$ real and $m$ virtual photons. Although the amplitudes of
these processes can be derived from corresponding Feynman diagrams
unambiguously, the length of the calculation formulae should
increase very fast and become very tedious even when $n+m=3$, let
alone the finial states containing more than three photons. As the
best to our knowledge, present calculations remain on the electron
positron annihilation into three real
photons~\cite{Berends:1980px} or two photons up to the next to
leading order~\cite{Rodrigo:2001jr}. In practice we utilize two
specific packages, FeynArts~\cite{Hahn:2000kx} and
FeynCalc~\cite{Mertig:1990an}, within the symbol calculation tool
Mathematica to do the programme work to overcome the complexity
problem we mentioned above and to avoid the potential mistakes
which may arise from the lengthy formulae. As our interest lies in
those particular processes whose final states involving virtual
photons, we introduce a new model package which contains
additional time-like massive photons obey the same dynamics of the
standard QED photons and should be considered as an extension to
the built-in QED model. The Feynman diagrams of a relatively
simple process $e^+e^-$ annihilate into one real photon and two
virtual photons are depicted in Fig.~\ref{fig:twophotonew} at tree
order as an example.

For a general process $e^-(p_1)e^+(p_2) \to
\gamma(k_1)\gamma(k_2)\gamma(k_3)$, the starting point is the
scattering formula \cite{wenberg}
\begin{equation}
 \mathrm{d}\sigma(\alpha\to\beta)\equiv \mathrm{d}\Gamma(\alpha \to
 \beta)/\Phi_\alpha = (2\pi)^4 u^{-1}_\alpha |M_{\beta\alpha}|^2
 \delta^4(p_\beta-p_\alpha) \mathrm{d}\beta
\label{eq:sigma}
\end{equation}
with
\begin{equation}
 u_\alpha \equiv \frac{|\vec{p}|(E_1+E_2)}{E_1 E_2} \;.
\label{eq:u}
\end{equation}
Here we get $u = 2$ when the electron and positron masses are
neglected. $d\beta$ is the 3-body phase space. The corresponding
amplitude
\begin{equation}
F\cdot M_{\beta\alpha}=\bar{v}(p_2)\slashed{\epsilon}^*(k_1)
\frac{\slashed{k}_1- \slashed{p}_2}{(k_1-p_2)^2}
\slashed{\epsilon}^*(k_2)
\frac{\slashed{k}_1+ \slashed{k}_2- \slashed{p}_2}{(k_1+k_2-p_2)^2}
 \slashed{\epsilon}^*(k_3)u(p_1) + \{\mathrm{perm}(k_1,k_2,k_3)\}
\label{eq:M}
\end{equation}
can be read from FeynArts directly with our modified-QED model,
where $u,v$ are spinors of initial states, $\epsilon$'s are
polarization vectors of final states, $p_i$ and $k_i$ are
corresponding momenta, $\mathrm{perm}(k_1,k_2,k_3)$ means all the
possible permutation of $k_1$, $k_2$, and $k_3$. And
$F=(2\pi)^{9/2}\sqrt{8E'_1E'_2E'_3}$ is a scalar factor arising from
the representation differences between FeynArts and FeynCalc
packages. Here we use the standard Lorentz covariant formula derived
from FeynArts instead of adopting the helicity form which is
proposed in Ref.~\cite{Kolodziej:1991pk} just for the technical
reason, because in Ref.~\cite{Kolodziej:1991pk} polarization
vectors, fermion spinors, and amplitude matrices are all expressed
in explicit forms. In order to speed up the numerical integral, we
derive a new phase space formula in Lorentz covariant form:
\begin{equation}
\delta^4(p_\beta-p_\alpha)\mathrm{d}\beta =
  \frac{|\vec{k}_1|^2 |\vec{k}_3| \sin\theta_1 \mathrm{d}\theta_1
    \mathrm{d}\theta_3 \mathrm{d}\phi_1 \mathrm{d}E'_1 \mathrm{d}E'_3
  }{\sin\theta_2 \sin(\phi_2-\phi_3)} =\frac{|\vec{k}_1|^2 |\vec{k}_3|
    \mathrm{d}\Omega_1\mathrm{d}\theta_3  \mathrm{d}E'_1 \mathrm{d}E'_3
  }{\sin\theta_2 \sin(\phi_2-\phi_3)} \;,
\label{eq:phase-space}
\end{equation}
where the sum over all possible solutions of energy-momentum
conservation respect to $\phi_2$ and $\phi_3$ is implied. The formula
is derived with the transformation of $\delta$ function:
\begin{equation}
\delta(f_1(\theta_1,\theta_2))\delta(f_2(\theta_1,\theta_2)) =
\sum\delta(\theta_1-\theta^0_1)\delta(\theta_2-\theta^0_2)
\left. A^{-1}\right|_{\theta_1=\theta^0_1,\theta_2=\theta^0_2} \;,
\end{equation}
where $f_1$ and $f_2$ are general functions depending on $\theta_1$ and
$\theta_2$,
\begin{equation}
A= \mathrm{det}\left| {\frac{\partial f_1}{\partial \theta_1} \quad
  \frac{\partial f_1}{\partial \theta_2} \atop \frac{\partial
    f_2}{\partial \theta_1} \quad \frac{\partial f_2}{\partial
    \theta_2}} \right|   \;,
\end{equation}
and $\theta^0_1$ and $\theta^0_2$ are the solutions of equations
$f_1(\theta_1,\theta_2)=0$ and $ f_2(\theta_1,\theta_2)=0$, $\sum$
means a sum over all possible solutions. Considering the
energy-momentum is conserved and the final state particles are
on-shell, we obtain Eq.~(\ref{eq:phase-space}). Combining
Eq.~(\ref{eq:sigma}) with Eqs.~(\ref{eq:u}),~(\ref{eq:M}),
and~(\ref{eq:phase-space}), we obtain the cross section formula
used in our program~\cite{footnote}.

\section{Numerical results}

\begin{figure}[hbtp]
\centering
\includegraphics[width=0.8\textwidth]{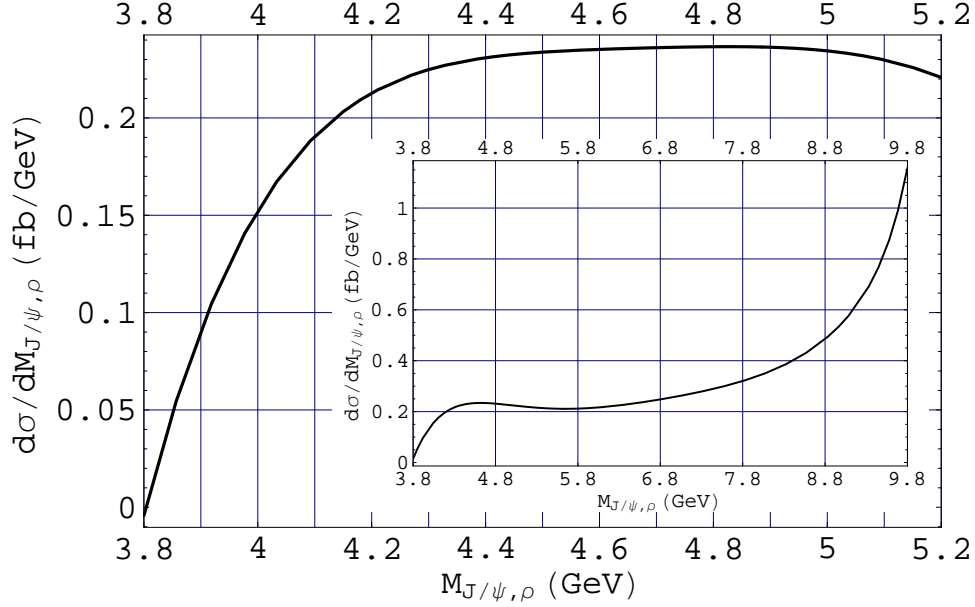}
\caption{Differential cross section of $e^+e^-\to \gamma
\gamma^*\gamma^* \to \gamma J/\psi \rho$ at $\sqrt{s}=10.58~\gev$,
the inset shows a wider mass range.} \label{fig:sigma0}
\end{figure}

\begin{figure}[hbtp]
\centering
\includegraphics[width=0.8\textwidth]{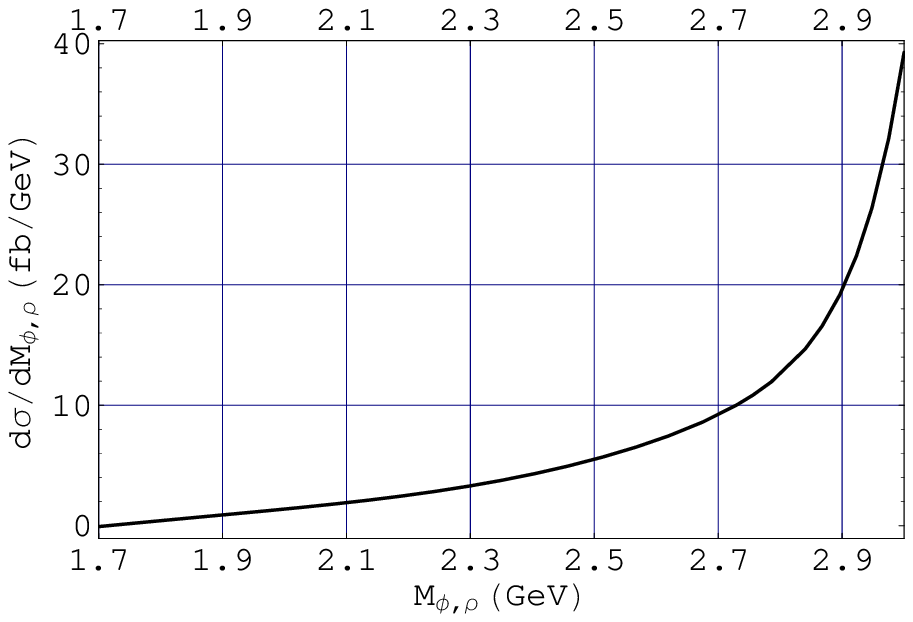}
\caption{Differential cross section of $e^+e^-\to \gamma
\gamma^*\gamma^* \to \gamma \phi\rho$ at $\sqrt{s}=3.097~\gev$.}
\label{fig:sigma2}
\end{figure}

\begin{figure}[hbtp]
\centering
\includegraphics[width=0.8\textwidth]{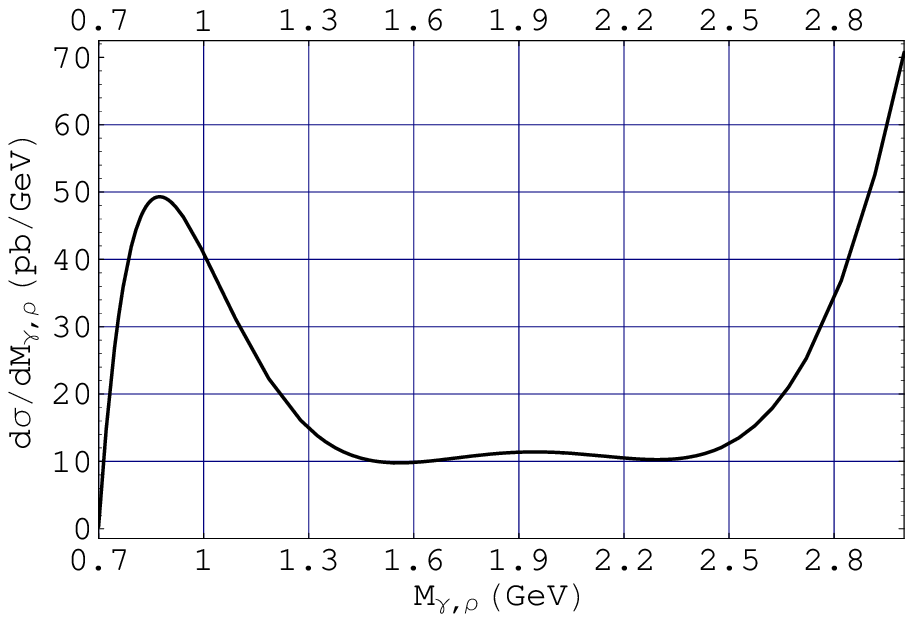}
\caption{Differential cross section of $e^+e^-\to \gamma
\gamma\gamma^* \to \gamma \gamma\rho$ at $\sqrt{s}=3.097~\gev$.}
\label{fig:sigma3}
\end{figure}

With the amplitudes of $e^+e^-\to \gamma \gamma^*\gamma^*$ and
$e^+e^-\to \gamma \gamma \gamma^*$, we obtain the cross sections
of $e^+e^-\to \gamma VV$ and $e^+e^-\to \gamma \gamma V$ by
Eq.~(\ref{eq:two-photon-model}). Since the final analytic results
are extremely lengthy, we only provide the numerical results here
for compactness. All the parameters used in our computation
without explicit exception are quoted from PDG~\cite{pdg:2006}. We
show the cross sections of processes $e^+e^-\to \gamma VV$ at
$\sqrt{s}=10.58~\gev$ in Table~\ref{tab:sigma1}, with $VV$ being
$J/\psi \rho$, $J/\psi \phi$ and $\rho\phi$ when the invariant
mass of the meson pair ranging from threshold to $5.2~\gev/c^2$.
As examples for illustration, the differential cross section of
$e^+e^-\to \gamma J/\psi\rho$ versus the invariant mass of $J/\psi
\rho$ is drawn in Fig.~\ref{fig:sigma0}. Similarly, the cross
sections of processes $e^+e^-\to \gamma VV$ at
$\sqrt{s}=3.097~\gev$ are exhibited in Table~\ref{tab:sigma2},
this time $VV$ represents $\phi\omega$ or $\phi\rho$, with the
invariant masses ranging from their thresholds to $3.0~\gev/c^2$.
Figure~\ref{fig:sigma2} shows the differential cross section of
$e^+e^-\to \gamma \phi\rho$ versus the invariant mass of
$\phi\rho$ as an example. Finally, we provide the cross sections
of $e^+e^-\to \gamma \gamma V$ at $\sqrt{s}=3.097~\gev$ in
Table~\ref{tab:sigma3}, here $V$ represents a single meson which
is $\rho$, $\phi$ or $\omega$, with the invariant mass of one
photon and one meson varying from their thresholds to
$3.0~\gev/c^2$. Here only the differential cross section of
$e^+e^-\to \gamma \gamma\rho$ versus the invariant mass of a
photon and $\rho$ is drawn in Fig.~\ref{fig:sigma3} as an example
for the same compactness reason.

\begin{table}[htbp]
\caption{{The cross section of $e^+e^-\to \gamma VV$ at
$\sqrt{s}=10.58~\gev$.}}
\begin{center}
\begin{tabular}{ccccc}
\hline\hline
final states &  & invariant mass of VV ($\gev/c^2$) &  & $\sigma~(\mathrm{fb})$  \\
\hline
$\gamma\ J/\psi \ \rho  $   &   &     $3.8-5.2$           &  & $0.29$ \\
\hline
$\gamma\  J/\psi\ \phi$  &   &     $4.1-5.2$           &  & $0.022$ \\
\hline
$\gamma\ \rho \ \phi $  &   &     $1.7-5.2$           &  & $0.66$ \\
\hline\hline
\end{tabular}
\end{center}
\label{tab:sigma1}
\end{table}

\begin{table}[htbp]
\caption{{The cross section of $e^+e^-\to \gamma VV$  at
 $\sqrt{s}=3.097~\gev$.}}
\begin{center}
\begin{tabular}{ccccc}
\hline\hline
final states &  & invariant mass of VV ($\gev/c^2$) &  & $\sigma~(\mathrm{fb})$  \\
\hline
$\gamma\ \phi \ \rho $  &   &     $1.7-3.0$           &  & $9.68$ \\
\hline
$\gamma\  \phi \ \omega $  &   &     $1.8-3.0$           &  & $0.80$ \\
\hline\hline
\end{tabular}
\end{center}
\label{tab:sigma2}
\end{table}

\begin{table}[htbp]
\caption{{The cross section of $e^+e^-\to \gamma \gamma V$ at
$\sqrt{s}=3.097~\gev$.}}
\begin{center}
\begin{tabular}{ccccc}
\hline\hline
final states &  & invariant mass of VV ($\gev/c^2$) &  & $\sigma~(\mathrm{pb})$  \\
\hline
$\gamma\ \gamma\ \rho $   &   &     $0.7-3.0$           &  & $49.30$ \\
\hline
$\gamma\ \gamma \ \phi $  &   &     $1.0-3.0$           &  & $6.75$ \\
\hline
$\gamma\ \gamma \ \omega $  &   &     $0.8-3.0$           &  & $3.96$ \\
\hline\hline
\end{tabular}
\end{center}
\label{tab:sigma3}
\end{table}

\section{Discussion}

In this article, we calculate the cross sections of $e^+e^-$
annihilation into one photon plus $J/\psi \rho$, $J/\psi \phi$, or
$\rho\phi$ at $\sqrt{s}=10.58~\gev$; and one photon plus
$\omega\phi$, $\rho\phi$, or $\gamma\rho$, $\gamma\phi$, or
$\gamma\omega$ at $\sqrt{s}=3.097~\gev$. The numerical results are
shown in Tables~\ref{tab:sigma1},~\ref{tab:sigma2},
and~\ref{tab:sigma3}, and several differential cross sections are
displayed as examples in Figs.~\ref{fig:sigma0},~\ref{fig:sigma2},
and~\ref{fig:sigma3}.

One essential component of our calculation is about the processes
$e^+e^-\to (n)\gamma(m)\gamma^*$, here we only calculate $e^+e^-\to
(1)\gamma(2)\gamma^*$ and $e^+e^-\to (2)\gamma(1)\gamma^*$ at
leading order. However, it should be pointed out that our method is
flexible and extensible, i.e., within the framework of our
computation, the annihilation is a purely QED process, any number of
photons and higher orders can be achieved without difficulty. For
any extension, the whole method is standard, only a few collateral
modifications need to be carried out within the computation frame.
More details are given in the following paragraph. We should also
mention that, up to now, we have neglected some possible corrections
from such as hadronic contributions, higher order loops, weak
interaction, product photons interference, and the mass
distributions of the resonances. However, these corrections are
small compared with the leading order contribution and far beyond
the present experimental precision. Similarly, the above improvement
can be achieved through just intuitive extensions of our present
calculation.

Here we take some interesting deductions from our calculation such
as $e^+e^- \to VV$ and $e^+e^-\to \gamma \gamma (\gamma)$. In
addition to illustrate the flexibility of our method, these
deductions also conform the validity of our computation by the
consistencies between them and other literatures. First, it is
obvious that when we reduce the number of photons from three to two,
we are actually calculating $e^+e^-$ annihilation into two virtual
photons. Then it is easy to get the cross sections with
two-vector-meson final states such as $e^+e^-\to \omega J/\psi$,
$e^+e^-\to \rho\phi$, and $e^+e^-\to \rho J/\psi$. It turns out that
the results from our calculation are the same as those in
Ref.~\cite{Davier:2006fu}. Note that the whole calculation is
standard, only a few specific considerations are taken into during
this reduction, such as the $2\to 2$ phase space and some "package
caused'' factors \cite{fnote} which are different from those in the
process $2\to 3$. Second, we can fix all the final state virtual
photons to be real, then actually we calculate the cross sections of
$e^+e^-\to \gamma\gamma\gamma$. Similar to the above case, here we
need only do a few peripheral modifications, that include fixing all
the photon-masses to zero since the photons are on-shell now,
setting the number of polarization directions to be two instead of
previous three, and multiplying a factor $1/3!=1/6$ because now the
final state contains just three identical bosons etc. Eventually our
result of $e^+e^-$ annihilation into three real photons is
consistent with both theoretical prediction~\cite{Berends:1980px}
and experimental result~\cite{Acciarri:1999cb}. Finally, with a
further reduction when we only calculate $e^+e^-$ annihilation into
two real photons, the analytic formula returns to the familiar
$(1+\cos^2\theta)/\sin^2\theta$ form. All the reductions we
discussed above are easily realized in our program and their
consistencies with other studies should be considered as
verifications of our method.

Finally we want to mention two important features of our results.
One is that if the phase space allowed, i.e. the CM energy is high
enough for a specific final state, all the differential cross
sections would show similar shapes. As what displayed in
Figs.~\ref{fig:sigma0} and~\ref{fig:sigma3}, with the invariant mass
varying from low to high, there would be a bump near the threshold
followed by a flat part and then ended with a fast increase when the
invariant mass approaches the CM energy, which is resulted from a
very soft radiated photon. The other feature is, as we expected,
these three-photon processes are $\mathcal{O}(\alpha)$ suppressed
compared with the corresponding $ISR$ processes where the hadron
system is produced from a single photon. That means the backgrounds
which arising directly from $e^+e^-\to \gamma V V$ and $e^+e^-\to
\gamma \gamma V$ are not essential at $B$-factories in the study of
the $ISR$ processes, nor at BES in $J/\psi$ decays at current
available statistics. However, accompanying with the upcoming
super-$B$ factory and BESIII, these modes will be important in the
near future when more accumulated luminosity is achieved and a
better precision is expected in various analyses.

\acknowledgments

This work is supported in part by the 100 Talents Program of CAS
under Contract No.~U-25 and by National Natural Science Foundation
of China under Contract No. 10491303.


\begin{thebibliography}{10}
\bibitem{fnote}For the process $2e\to 2\gamma$, the scalar
factor is $F=(2\pi)^3\sqrt{4E'_1E'_2}$.
\bibitem{footnote} Actually we also need to average initial state and
sum over all particular polarizations of the final state particles.
In FeynCalc, for an arbitrary $4\times4$ matrix $A$, we use a
built-in function FermionSpinSum[~] to construct the Traces out the
squared amplitude $\sum_{\sigma\sigma'}|\bar{u}(p',\sigma')
Au(p,\sigma)|^2 = \mathrm{Tr}\left\{ A
\left(\frac{-i\slashed{p}+m}{2p^0}\right) \beta A^\dag \beta
\left(\frac{-i\slashed{p}'+m}{2p'^0}\right)\right\}$, and we use
another built-in function PolarizationSum[~] to reduce the
polarization sum: for real and virtual photons they are
$\epsilon^*_\mu(k)\epsilon_\nu(k)=-g_{\mu\nu}$ and
$\epsilon^*_\mu(k)\epsilon_\nu(k)=k_\mu k_\nu/k^2 -g_{\mu\nu}$
respectively.
\end{thebibliography}
\end{document}